# Deep Learning in Alzheimer's Disease: Diagnostic Classification and Prognostic Prediction using Neuroimaging Data

**Taeho Jo[1-3]\*, Kwangsik Nho[1-3], Andrew J. Saykin[1-3]**

[1]Center for Neuroimaging, Department of Radiology and Imaging Sciences, Indiana University School of Medicine, Indianapolis, IN, USA

[2]Indiana Alzheimer Disease Center, Indiana University School of Medicine, Indianapolis, IN, USA

[3]Indiana University Network Science Institute, Bloomington, IN, USA

**\*Correspondence:**

Taeho Jo, PhD

Department of Radiology and Imaging Sciences

Indiana University School of Medicine

Indianapolis, IN, USA

Phone: 317-963-7505; Fax: 317-274-1067; E-mail: tjo@iu.edu

**Abstract**

Deep learning, a state-of-the-art machine learning approach, has shown outstanding performance over traditional machine learning in identifying intricate structures in complex high-dimensional data, especially in the domain of computer vision. The application of deep learning to early detection and automated classification of Alzheimer's disease (AD) has recently gained considerable attention, as rapid progress in neuroimaging techniques has generated large-scale multimodal neuroimaging data. A systematic review of publications using deep learning approaches and neuroimaging data for diagnostic classification of AD was performed. A PubMed and Google Scholar search was used to identify deep learning papers on AD published between January 2013 and July 2018. These papers were reviewed, evaluated, and classified by algorithm and neuroimaging type, and the findings were summarized. Of 16 studies meeting full inclusion criteria, 4 used a combination of deep learning and traditional machine learning approaches, and 12 used only deep learning approaches. The combination of traditional machine learning for classification and stacked auto-encoder (SAE) for feature selection produced accuracies of up to 98.8% for AD classification and 83.7% for prediction of conversion from mild cognitive impairment (MCI), a prodromal stage of AD, to AD. Deep learning approaches, such as convolutional neural network (CNN) or recurrent neural network (RNN), that use neuroimaging data without preprocessing for feature selection have yielded accuracies of up to 96.0% for AD classification and 84.2% for MCI conversion prediction. The best classification performance was obtained when multimodal neuroimaging and fluid biomarkers were combined. Deep learning approaches continue to improve in performance and appear to hold promise for diagnostic classification of AD using multimodal neuroimaging data. AD research that uses deep learning is still evolving, improving performance by incorporating additional hybrid data types, such as –omics data, increasing transparency with explainable approaches that add knowledge of specific disease-related features and mechanisms.

# 1 Introduction

Alzheimer's disease (AD), the most common form of dementia, is a major challenge for healthcare in the 21$^{st}$ century. An estimated 5.5 million people aged 65 and older are living with AD, and AD is the sixth-leading cause of death in the United States. The global cost of managing AD, including medical, social welfare, and salary loss to the patients' families, was $277 billion in 2018 in the United States, heavily impacting the overall economy and stressing the U.S. health care system. (Alzheimer's Association, 2018). AD is an irreversible, progressive brain disorder marked by a decline in cognitive functioning with no validated disease modifying treatment (De Strooper and Karran, 2016). Thus, a great deal of effort has been made to develop strategies for early detection, especially at pre-symptomatic stages in order to slow or prevent disease progression (Galvin, 2017;Schelke et al., 2018). In particular, advanced neuroimaging techniques, such as magnetic resonance imaging (MRI) and positron emission tomography (PET), have been developed and used to identify AD-related structural and molecular biomarkers (Veitch et al., 2019). Rapid progress in neuroimaging techniques has made it challenging to integrate large-scale, high dimensional multimodal neuroimaging data. Therefore, interest has grown rapidly in computer-aided machine learning approaches for integrative analysis. Well-known pattern analysis methods, such as linear discriminant analysis (LDA), linear program boosting method (LPBM), logistic regression (LR), support vector machine (SVM), and support vector machine-recursive feature elimination (SVM-RFE), have been used and hold promise for early detection of AD and the prediction of AD progression (Rathore et al., 2017).

In order to apply such machine learning algorithms, appropriate architectural design or pre-processing steps must be predefined (Lu and Weng, 2007). Classification studies using machine learning generally require four steps: feature extraction, feature selection, dimensionality reduction, and feature-based classification algorithm selection. These procedures require specialized knowledge and multiple stages of optimization, which may be time-consuming. Reproducibility of these approaches has been an issue (Samper-Gonzalez et al., 2018). For example, in the feature selection process, AD-related features are chosen from various neuroimaging modalities to derive more informative combinatorial measures, which may include mean subcortical volumes, gray matter densities, cortical thickness, brain glucose metabolism, and cerebral amyloid-β accumulation in regions of interest (ROIs), such as the hippocampus (Riedel et al., 2018).

In order to overcome these difficulties, deep learning, an emerging area of machine learning research that uses raw neuroimaging data to generate features through "on-the-fly" learning, is attracting considerable attention in the field of large-scale, high-dimensional medical imaging analysis (Plis et al., 2014). Deep learning methods, such as convolutional neural networks (CNN), have been shown to outperform existing machine learning methods (LeCun et al., 2015).

We systematically reviewed publications where deep learning approaches and neuroimaging data were used for the early detection of AD and the prediction of AD progression. A PubMed and Google Scholar search was used to identify deep learning papers on AD published between January 2013 and July 2018. The papers were reviewed and evaluated, classified by algorithms and neuroimaging types, and the findings were summarized. In addition, we discuss challenges and implications for the application of deep learning to AD research.

# 2 Deep Learning Methods

Deep learning is a subset of machine learning (LeCun et al., 2015), meaning that it learns features through a hierarchical learning process (Bengio, 2009). Deep learning methods for classification or prediction have been applied in various fields, including computer vision (Ciregan et al., 2012;Krizhevsky et al., 2012;Farabet et al., 2013) and natural language processing (Hinton et al., 2012;Mikolov et al., 2013), both of which demonstrate breakthroughs in performance (Boureau et al., 2010;Russakovsky et al., 2015). Because deep learning methods have been reviewed extensively in recent years (Bengio, 2013;Bengio et al., 2013;Schmidhuber, 2015), we focus here on basic concepts of Artificial Neural Networks (ANN) that underlie deep learning (Hinton and Salakhutdinov, 2006). We also discuss architectural layouts of deep learning that have been applied to the task of AD classification and prognostic prediction. ANN is a network of interconnected processing units called artificial neurons that were modeled (McCulloch and Pitts, 1943) and developed with the concept of Perceptron (Rosenblatt, 1957;1958), Group Method of Data Handling (GMDH) (Ivakhnenko and Lapa, 1965;Ivakhnenko, 1968;1971) and the Neocognitron (Fukushima, 1979;1980). Efficient error functions and gradient computing methods were discussed in these seminal publications, spurred by the demonstrated limitation of the single layer perceptron, which can learn only linearly separable patterns (Minsky and Papert, 1969). Further, the back-propagation procedure, which uses gradient descent, was developed and applied to minimize the error function (Werbos, 1982;Rumelhart et al., 1986;LeCun et al., 1988;Werbos, 2006).

## 2.1 Gradient Computation

The back-propagation procedure is used to calculate the error between the network output and the expected output. The back propagation calculates the gap repeatedly, changing weights and stopping the calculation when the gap is no longer updated. (Rumelhart et al., 1986;Bishop, 1995;Ripley and Hjort, 1996;Schalkoff, 1997). Figure 1 illustrates the process of the neural network made by multilayer perceptron. After the initial error value is calculated from the given random weight by the least squares method, the weights are updated until the differential value becomes 0. For example, the $w_{31}$ in Figure 1 is updated by the following formula:

$$w_{31}(t+1) = w_{31}t - \frac{\partial ErrorY_{out}}{\partial w_{31}}$$

$$ErrorY_{out} = \frac{1}{2}(y_{t1} - y_{o1})^2 + \frac{1}{2}(y_{t2} - y_{o2})^2$$

The $ErrorY_{out}$ is the sum of error $y_{o1}$ and error $y_{o2}$. $y_{t1}$, $y_{t2}$ are constants that are known through the given data. The partial derivative of $ErrorY_{out}$ with respect to $w_{31}$ can be calculated by the chain rule as follows.

$$\frac{\partial ErrorY_{out}}{\partial w_{31}} = \frac{\partial ErrorY_{out}}{\partial y_{o1}} \cdot \frac{\partial y_{o1}}{\partial net3} \cdot \frac{\partial net3}{\partial w_{31}}$$

Likewise, $w_{11}$ in the hidden layer is updated by the chain rule as follows.

$$\frac{\partial ErrorY_{out}}{\partial w_{11}} = \frac{\partial ErrorY_{out}}{\partial y_{h1}} \cdot \frac{\partial y_{h1}}{\partial net_1 y} \cdot \frac{\partial net_1}{\partial w_{11}}$$

Detailed calculation of the weights in the backpropagation is described in **Supplement 1**.

## 2.2 Modern Practical Deep Neural Networks

As the back-propagation uses a gradient descent method to calculate the weights of each layer going backwards from the output layer, a vanishing gradient problem occurs as the layer is stacked, where the differential value becomes 0 before finding the optimum value. As shown in Figure 2a, when the sigmoid is differentiated, the maximum value is 0.25, which becomes closer to 0 when it continues to multiply. This is called a vanishing gradient issue, a major obstacle of the deep neural network. Considerable research has addressed the challenge of the vanishing gradient (Goodfellow et al., 2016). One of the accomplishments of such an effort is to replace the sigmoid function, an activation function, with several other functions, such as the hyperbolic tangent function, ReLu and Softplus (Nair and Hinton, 2010;Glorot et al., 2011). The hyperbolic tangent (tanh) function expands the range of derivative values of the sigmoid. The ReLu function, the most used activation function, replaces a value with 0 when the value is less than 0 and uses the value if the value is greater than 0. As the derivative becomes 1, when the value is larger than 0, it becomes possible to adjust the weights without disappearing up to the first layer through the stacked hidden layers. This simple method allows building multiple layers and accelerates the development of deep learning. The Softplus function replaces the ReLu function with a gradual descent method when ReLu becomes zero.

While a gradient descent method is used to calculate the weights accurately, it usually requires a large amount of computation time because all of the data needs to be differentiated at each update. Thus, in addition to the activation function, advanced gradient descent methods have been developed to solve speed and accuracy issues. For example, Stochastic Gradient Descent (SGD) uses a subset that is randomly extracted from the entire data for faster and more frequent updates (Bottou, 2010), and it has been extended to Momentum SGD (Sutskever et al., 2013). Currently, one of the most popular gradient descent method is Adaptive Moment Estimation (Adam). Detailed calculation of the optimization methods is described in **Supplement 2**.

## 2.3 Architectures of Deep Learning

Overfitting has also played a major role in the history of deep learning (Schmidhuber, 2015), with efforts being made to solve it at the architectural level. The Restricted Boltzmann Machine (RBM) was one of the first models developed to overcome the overfitting problem (Hinton and Salakhutdinov, 2006). Stacking the RBMs resulted in building deeper structures known as the Deep Boltzmann Machine (DBM) (Salakhutdinov and Larochelle, 2010). The Deep Belief Network (DBN) is a supervised learning method used to connect unsupervised features by extracting data from each stacked layer (Hinton et al., 2006). DBN was found to have a superior performance to other models and is one of the reasons that deep learning has gained popularity (Bengio, 2009). While DBN solves the overfitting problem by reducing the weight initialization using RBM, CNN efficiently reduces the number of model parameters by inserting convolution and pooling layers that lead to a reduction in complexity. Because of its effectiveness, when given enough data, CNN is widely used in the field of visual recognition. Figure 3 shows the structures of RBM, DBM, DBN, CNN, Auto-Encoders (AE), sparse AE, and stacked AE respectively. Auto-Encoders (AE) are an unsupervised learning method that make the output value approximate to the input value by using the back-propagation and SGD (Hinton and Zemel, 1994). AE engages the dimensional reduction, but it is difficult to train due to the vanishing gradient issue. Sparse AE has solved this issue by allowing for only a small number of the hidden units (Makhzani and Frey, 2013). Stacked AE stacks sparse AE like DBN.

DNN, RBM, DBM, DBN, AE, Sparse AE, and Stacked AE are deep learning methods that have been used for Alzheimer's disease diagnostic classification to date (see Table 1 for the definition of acronyms). Each approach has been developed to classify AD patients from cognitively normal controls (CN) or mild cognitive impairment (MCI), which is the prodromal stage of AD. Each approach is used to predict the conversion of MCI to AD using multi-modal neuroimaging data. In this paper, when deep learning is used together with traditional machine learning methods, i.e., SVM as a classifier, it is referred to as a 'hybrid method'.

## 3    Materials and methods

We conducted a systematic review on previous studies that used deep learning approaches for diagnostic classification of AD with multimodal neuroimaging data. The search strategy is outlined in detail using the PRISMA flow diagram (Moher et al., 2009) in Figure 4.

### 3. 1 Identification

From a total of 389 hits on Google scholar and PubMed search, 16 articles were included in the systematic review.

Google Scholar: We searched using the following key words and yielded 358 results. ("Alzheimer disease" OR "Alzheimer's disease"), ("deep learning" OR "deep neural network" OR "convolutional neural network" OR "CNN" OR "Autoencoder" OR "Deep Belief Network" OR "Restricted Boltzmann Machine"),("Neuroimaging" OR "MRI" OR "multimodal")

PubMed: The keywords used in the Google Scholar search were reused for the search in PubMed, and yielded 31 search results. ("Alzheimer disease" OR "Alzheimer's disease") AND ("deep learning" OR "deep neural network" OR "convolutional neural network" OR "recurrent neural network" OR "Auto-Encoder" OR "Auto Encoder" OR "Restricted Boltzmann Machine" OR "Deep Belief Network" OR "Generative Adversarial Network" OR "Reinforcement Learning" OR "Long Short Term Memory" OR "Gated Recurrent Units")AND ("Neuroimaging" OR "Magnetic Resonance Imaging" OR "multimodal")

Among the 389 relevant records, 25 overlapping records were removed.

### 3.2 Screening based on article type

We first excluded 38 survey papers, 22 theses, 19 Preprint, 34 book chapters, 20 conference abstract, 13 none English papers, 5 citations and 10 patents. We also excluded 11 papers of which the full text was not accessible. The remaining 192 articles were downloaded for review.

### 3.3 Eligibility screening

Out of the 192 publications retrieved, 150 articles were excluded because the authors only introduced or mentioned deep learning but did not use it. Out of the 42 remaining publications, (1) 18 articles were excluded because they did not perform deep learning approaches for AD classification and/or prediction of MCI to AD conversion; (2) 5 articles were excluded because their neuroimaging data were not explicitly described; and (3) 3 articles were excluded because performance results were not provided. The remaining 16 papers were included in this review for AD classification and/or prediction of MCI to AD conversion. All of the final selected and compared papers used ADNI data in common.

## 4 Results

From the 16 papers included in this review, Table 2 provides the top results of diagnostic classification and/or prediction of MCI to AD conversion. We compared only binary classification results. Accuracy is a measure used consistently in the sixteen publications. However, it is only one metric of the performance characteristics of an algorithm. The group composition, sample sizes, and number of scans analyzed are also noted together because accuracy is sensitive to unbalanced distributions. Table S1 shows the full results sorted according to the performance accuracy as well as the number of subjects, the deep learning approach, and the neuroimaging type used in each paper.

### 4.1 Deep learning for feature selection from neuroimaging data

Multimodal neuroimaging data have been used to identify structural and molecular/functional biomarkers for AD. It has been shown that volumes or cortical thicknesses in pre-selected AD-specific regions, such as the hippocampus and entorhinal cortex, could be used as features to enhance the classification accuracy in machine learning. Deep learning approaches have been used to select features from neuroimaging data.

As shown in Figure 5, 4 studies have used hybrid methods that combine deep learning for feature selection from neuroimaging data and traditional machine learning, such as the SVM as a classifier. Suk and Shen (2013) used a stacked auto-encoder (SAE) to construct an augmented feature vector by concatenating the original features with outputs of the top hidden layer of the representative SAEs. Then, they used a multi-kernel SVM for classification to show 95.9% accuracy for AD/CN classification and 75.8% prediction accuracy of MCI to AD conversion. These methods successfully tuned the input data for the SVM classifier. However, SAE as a classifier (Suk et al., 2015) yielded 89.9% accuracy for AD/CN classification and 60.2% accuracy for prediction of MCI to AD conversion. Later Suk et al. (2015) extended the work to develop a two-step learning scheme: greedy layer-wise pre-training and fine-tuning in deep learning. The same authors further extended their work to use the DBM to find latent hierarchical feature representations by combining heterogeneous modalities during the feature representation learning (Suk et al., 2014). They obtained 95.35% accuracy for AD/CN classification and 74.58% prediction accuracy of MCI to AD conversion. In addition, the authors initialized SAE parameters with target-unrelated samples and tuned the optimal parameters with target-related samples to have 98.8% accuracy for AD/CN classification and 83.7% accuracy for prediction of MCI to AD conversion (Suk et al., 2015). Li et al. (2015) used the RBM with a dropout technique to reduce overfitting in deep learning and SVM as a classifier, which produced 91.4% accuracy for AD/CN classification and 57.4% prediction accuracy of MCI to AD conversion.

### 4.2 Deep learning for diagnostic classification and prognostic prediction

To select optimal features from multimodal neuroimaging data for diagnostic classification, we usually need several pre-processing steps, such as neuroimaging registration and feature extraction, which greatly affect the classification performance. However, deep learning approaches have been applied to AD diagnostic classification using original neuroimaging data without any feature selection procedures.

As shown in Figure 5, 12 studies have used only deep learning for diagnostic classification and/or prediction of MCI to AD conversion. Liu et al. (2014) used stacked sparse auto-encoders (SAEs) and a softmax regression layer and showed 87.8% accuracy for AD/CN classification. Liu et al. (2015) used SAE and a softmax logistic regressor as well as a zero-mask strategy for data fusion to extract complementary information from multimodal neuroimaging data (Ngiam et al., 2011), where one of the modalities is randomly hidden by replacing the input values with zero to converge different types of image data for SAE. Here, the deep learning algorithm improved accuracy for AD/CN classification by 91.4%. Recently, Lu et al. (2018) used SAE for pre-training and DNN in the last step, which achieved an AD/CN classification accuracy of 84.6% and an MCI conversion prediction accuracy of 82.93%. CNN, which has shown remarkable performance in the field of image recognition, has also been used for the diagnostic classification of AD with multimodal neuroimaging data. Cheng et al. (2017) used image patches to transform the local images into high-level features from the original MRI images for the 3D-CNN and yielded 87.2% accuracy for AD/CN classification. They improved the accuracy to 89.6% by running two 3D-CNNs on neuroimage patches extracted from MRI and PET separately and by combining their results to run 2D CNN (Cheng and Liu, 2017). Korolev et al. (2017) applied two different 3D CNN approaches (plain (VoxCNN) and residual neural networks (ResNet)) and reported 80% accuracy for AD/CN classification, which was the first study that the manual feature extraction step was unnecessary. Aderghal et al. (2017) captured 2D slices from the hippocampal region in the axial, sagittal, and coronal directions and applied 2D CNN to show 85.9% accuracy for AD/CN classification. Liu et al. (2018b) selected discriminative patches from MR images based on AD-related anatomical landmarks identified by a data-driven learning approach and ran 3D CNN on them. This approach used three independent data sets (ADNI-1 as training, ADNI-2 and MIRIAD as testing) to yield relatively high accuracies of 91.09% and 92.75% for AD/CN classification from ADNI-2 and MIRIAD, respectively, and an MCI conversion prediction accuracy of 76.9% from ADNI-2. Li et al. (2014) trained 3D CNN models on subjects with both MRI and PET scans to encode the nonlinear relationship between MRI and PET images and then used the trained network to estimate the PET patterns for subjects with only MRI data. This study obtained an AD/CN classification accuracy of 92.87% and an MCI conversion prediction accuracy of 72.44%. Vu et al. (2017) applied SAE and 3D CNN to subjects with MRI and FDG PET scans to yield an AD/CN classification accuracy of 91.1%. Liu et al. (2018a) decomposed 3D PET images into a sequence of 2D slices and used a combination of 2D CNN and RNNs to learn the intra-slice and inter-slice features for classification, respectively. The approach yielded AD/CN classification accuracy of 91.2%. If the data is imbalanced, the chance of misdiagnosis increases and sensitivity decreases. For example, in Suk et al. (2014) there were 76 cMCI and 128 ncMCI subjects and the obtained sensitivity of 48.04% was low. Similarly, Liu et al. (2018b) included 38 cMCI and 239 ncMCI subjects and had a low sensitivity of 42.11%. Recently Choi and Jin (2018) reported the first use of 3D CNN models to multimodal PET images (FDG PET and [18F]florbetapir PET) and obtained 96.0% accuracy for AD/CN classification and 84.2% accuracy for the prediction of MCI to AD conversion.

### 4.3  Performance comparison by types of neuroimaging techniques

In order to improve the performance for AD/CN classification and for the prediction of MCI to AD conversion, multimodal neuroimaging data such as MRI and PET have commonly been used in deep learning: MRI for brain structural atrophy, amyloid PET for brain amyloid-β accumulation, and FDG-PET for brain glucose metabolism. MRI scans were used in 13 studies, FDG-PET scans in 10, both MRI and FDG-PET scans in 12, and both amyloid PET and FDG-PET scans in 1. The performance in AD/CN classification and/or prediction of MCI to AD conversion yielded better results in PET data compared to MRI. Two or more multimodal neuroimaging data types produced higher accuracies than a single neuroimaging technique. Figure 6 shows the results of the performance comparison by types of neuroimaging techniques.

**4.4 Performance comparison by deep learning algorithms**

Deep learning approaches require massive amounts of data to achieve the desired levels of performance accuracy. In currently limited neuroimaging data, the hybrid methods that combine traditional machine learning methods for diagnostic classification with deep learning approaches for feature extraction yielded better performance and can be a good alternative to handle the limited data. Here, an auto-encoder (AE) was used to decode the original image values, making them similar to the original image, which it then included as input, thereby effectively utilizing the limited neuroimaging data. Although hybrid approaches have yielded relatively good results, they do not take full advantage of deep learning, which automatically extracts features from large amounts of neuroimaging data. The most commonly used deep learning method in computer vision studies is the CNN, which specializes in extracting characteristics from images. Recently, 3D CNN models using multimodal PET images (FDG-PET and [18F]florbetapir PET) showed better performance for AD/CN classification and for the prediction of MCI to AD conversion.

**5 Discussion**

Effective and accurate diagnosis of Alzheimer's disease (AD) is important for initiation of effective treatment. Particularly, early diagnosis of AD plays a significant role in therapeutic development and ultimately for effective patient care. In this study, we performed a systematic review of deep learning approaches based on neuroimaging data for diagnostic classification of AD. We analyzed 16 articles published between 2013 and 2018 and classified them according to deep learning algorithms and neuroimaging types. Among 16 papers, 4 studies used a hybrid method to combine deep learning and traditional machine learning approaches as a classifier, and 12 studies used only deep learning approaches. In a limited available neuroimaging data set, hybrid methods have produced accuracies of up to 98.8% for AD classification and 83.7% for prediction of conversion from MCI to AD. Deep learning approaches have yielded accuracies of up to 96.0% for AD classification and 84.2% for MCI conversion prediction. While it is a source of concern when experiments obtain a high accuracy using small amounts of data, especially if the method is vulnerable to overfitting, the highest accuracy of 98.8% was due to the SAE procedure, whereas the 96% accuracy was due to the amyloid PET scan, which included pathophysiological information regarding AD. The highest accuracy for the AD classification was 87% when 3DCNN was applied from the MRI without the feature extraction step (Cheng et al., 2017). Therefore, two or more multimodal neuroimaging data types have been shown to produce higher accuracies than a single neuroimaging type.

In traditional machine learning, well-defined features influence performance results. However, the greater the complexity of the data, the more difficult it is to select optimal features. Deep learning identifies optimal features automatically from the data (i.e., the classifier trained by deep learning finds features that have an impact on diagnostic classification without human intervention). Because of its ease-of-use and better performance, deep learning has been used increasingly for medical image analysis. The number of studies of AD using CNN, which show better performance in image recognition among deep learning algorithms, has increased drastically since 2015. This is consistent with a previous survey showing that the use of deep learning for lesion classification, detection and segmentation has also increased rapidly since 2015 (Litjens et al., 2017).

Recent trends in the use of deep learning are aimed at faster analysis with better accuracy than human practitioners. Google's well-known study for the diagnostic classification of diabetic retinopathy (Gulshan et al., 2016) showed classification performance that goes well beyond that of a skilled professional. The diagnostic classification by deep learning needs to show consistent performance under various conditions, and the predicted classifier should be interpretable. In order for diagnostic classification and prognostic prediction using deep learning to reach readiness for real world clinical applicability, several issues need to be addressed, as discuss below.

## 5.1 Transparency

Traditional machine learning approaches may require expert involvement in preprocessing steps for feature extraction and selection from images. However, since deep learning does not require human intervention but instead extracts features directly from the input images, the data preprocessing procedure is not routinely necessary, allowing flexibility in the extraction of properties based on various data-driven inputs. Therefore, deep learning can create a good, qualified model at each time of the run. The flexibility has shown deep learning to achieve a better performance than other traditional machine learning that relies on preprocessing (Bengio, 2013). However, this aspect of deep learning necessarily brings uncertainty over which features would be extracted at every epoch, and unless there is a special design for the feature, it is very difficult to show which specific features were extracted within the networks (Goodfellow et al., 2016). Due to the complexity of the deep learning algorithm, which has multiple hidden layers, it is also difficult to determine how those selected features lead to a conclusion and to the relative importance of specific features or subclasses of features. This is a major limitation for mechanistic studies where understanding the informativeness of specific features is desirable for model building. These uncertainties and complexities tend to make the process of achieving high accuracy opaque and also make it more difficult to correct any biases that arise from a given data set. This lack of clarity also limits the applicability of obtained results to other use cases.

The issue of transparency is linked to the clarity of the results from machine learning and is not a problem limited to deep learning (Kononenko, 2001). Despite the simple principle, the complexity of the algorithm makes it difficult to describe mathematically. When one perceptron advances to a neural network by adding more hidden layers, it becomes even more difficult to explain why a particular prediction was made. AD classification based on 3D multimodal medical images with deep learning involves nonlinear convolutional layers and pooling that have different dimensionality from the source data, making it very difficult to interpret the relative importance of discriminating features in original data space. This is a fundamental challenge in view of the importance of anatomy in the interpretation of medical images, such as MRI or PET scans. The more advanced algorithm generates plausible results, but the mathematical background is difficult to explain, although the output for diagnostic classification should be clear and understandable.

## 5.2 Reproducibility

Deep learning performance is sensitive to the random numbers generated at the start of training, and hyper-parameters, such as learning rates, batch sizes, weight decay, momentum, and dropout probabilities, may be tuned by practitioners (Hutson, 2018). To produce the same experimental result, it is important to set the same random seeds on multiple levels. It is also important to maintain the same code bases (Vaswani et al., 2018), even though the hyper-parameters and random seeds were not, in most cases, provided in our study. The uncertainty of the configuration and the randomness involved in the training procedure may make it difficult to reproduce the study and achieve the same results.

When the available neuroimaging data is limited, careful consideration at the architectural level is needed to avoid the issues of overfitting and reproducibility. Data leakage in machine learning (Smialowski et al., 2009) occurs when the data set framework is designed incorrectly, resulting in a model that uses inessential additional information for classification. In the case of diagnostic classification for the progressive and irreversible Alzheimer's disease, all subsequent MRI images should be labeled as belonging to a patient with Alzheimer's disease. Once the brain structure of the patient is shared by both the training and testing sets, the morphological features of the patient's brain greatly influence the classification decision, rather than the biomarkers of dementia. In the present study, articles were excluded from the review if the data set configurations did not explicitly describe how to prevent data leakage.

Future studies ultimately need to replicate key findings from deep learning on entirely independent data sets. This is now widely recognized in genetics (König, 2011;Bush and Moore, 2012) and other fields but has been slow to penetrate deep learning studies employing neuroimaging data. Hopefully the emerging open ecology of medical research data, especially in the AD and related disorders field (Toga et al., 2016;Reas, 2018), will provide a basis to remediate this problem.

## 6  Outlook and Future direction

Deep Learning algorithms and applications continue to evolve, producing the best performance in closed-ended cases, such as image recognition (Marcus, 2018). It works particularly well when inference is valid, i.e., the training and test environments are similar. This is especially true in the study of AD when using neuroimages (Litjens et al., 2017). One weakness of deep learning is that it is difficult to modify potential bias in the network when the complexity is too great to guarantee transparency and reproducibility. The issue may be solved through the accumulation of large-scale neuroimaging data and by studying the relationships between deep learning and features. Disclosing the parameters used to obtain the results and mean values from sufficient experimentations can mitigate the issue of reproducibility.

Not all problems can be solved with deep learning. Deep learning that extracts attributes directly from the input data without preprocessing for feature selection has difficulty integrating different formats of data as an input, such as neuroimaging and genetic data. Because the adjustment of weights for the input data is performed automatically within a closed network, adding additional input data into the closed network causes confusion and ambiguity. A hybrid approach, however, puts the additional information into machine learning parts and the neuroimages into deep learning parts before combining the two results.

Progress will be made in deep learning by overcoming these issues while presenting problem-specific solutions. As more and more data are acquired, research using deep learning will become more impactful. The expansion of 2D CNN into 3D CNN is important, especially in the study of AD, which deals with multimodal neuroimages. In addition, Generative Adversarial Networks (GAN) (Goodfellow et al., 2014) may be applicable for generating synthetic medical images for data augmentation. Furthermore, reinforcement learning (Sutton and Barto, 2018), a form of learning that adapts to changes in data as it makes its own decision based on the environment, may also demonstrate applicability in the field of medicine.

AD research using deep learning is still evolving to achieve better performance and transparency. As multimodal neuroimaging data and computer resources grow rapidly, research on the diagnostic classification of AD using deep learning is shifting towards a model that uses only deep learning algorithms rather than hybrid methods, although methods need to be developed to integrate completely different formats of data in a deep learning network.

## 7 Conflict of Interest

*The authors declare that the research was conducted in the absence of any commercial or financial relationships that could be construed as a potential conflict of interest.*

## 8 Author Contributions

Conceptualization and study design (TJ, AJS), Data collection and analysis (TJ), Drafting manuscript (TJ), Revision of the manuscript for important scientific content (TJ, KN, AJS), Final approval (TJ, KN, AJS).

## 9 Funding

This review was supported, in part, by grants from the National Institutes of Health (NIH) and include the following sources: P30 AG10133, R01 AG19771, R01 AG057739, R01 CA129769, NLM R01 LM012535, and NIA R03 AG054936. Many studies reviewed here analyzed data from the Alzheimer's Disease Neuroimaging Initiative (ADNI) that was funded by the National Institutes of Health (U01 AG024904) and Department of Defense (W81XWH-12-2-0012) and a consortium of private partners.

## 10 Acknowledgments

We are grateful to all of the study participants and their families that participated in the neuroimaging research on Alzheimer's disease reviewed here. We are also indebted to the clinical and computational researchers who reported their results, facilitating the analyses and discussion in this systematic review. We thank Paula J. Bice, Ph.D., for editorial assistance.

# 11 Tables

**Table1.** Definition of acronyms

| Acronym | Description | Acronym | Description |
|---|---|---|---|
| **ANN** | Artificial Neural Network | **CNN** | Convolutional Neural Network |
| **DNN** | Deep Neural Network | **RNN** | Recurrent Neural Network |
| **RBM** | Restricted Boltzmann Machine | **GAN** | Generative Adversarial Networks |
| **DBM** | Deep Boltzmann Machine | **SGD** | Stochastic Gradient Descent |
| **DBN** | Deep Belief Network | **SVM** | Support Vector Machine |
| **AE** | Auto-Encoders | **ROI** | Regions of Interest |
| **SAE** | Stacked Auto-Encoder | **HMM** | Hidden Markov Model |

Table2. Summary of 16 previous studies to systematically be reviewed

| Author (year) | Modality | Data processing / training | Classifier | AD:NC acc. | SEN | SPE | cMCI:ncMCI acc. | SEN | SPE | AD | cMCI | ncMCI | NC | Total |
|---|---|---|---|---|---|---|---|---|---|---|---|---|---|---|
| **Suk and Shen (2013)** | MRI,PET,CSF | SAE | SVM | **95.9** | | | **75.8** | | | 51 | 43 | 56 | 52 | 202 |
| **Liu et al. (2014)** | MRI,PET | SAE + NN | softmax | **87.76** | 88.57 | 87.22 | 76.92(MCI:NC) | 74.29 | 78.13 | 65 | 67 | 102 | 77 | 311 |
| **Suk et al. (2014)** | MRI,PET | DBM | SVM | **95.35** | 94.65 | 95.22 | **75.92** 85.67(MCI:NC) | 48.04 95.37 | 95.23 65.87 | 93 | 76 | 128 | 101 | 398 |
| **Li et al. (2014)** | MRI,PET | 3D CNN | Logistic regression | **92.87** | | | 76.21(MCI:NC) | | | 198 | 167 | 236 | 229 | 830 |
| **Li et al. (2015)** | MRI,PET,CSF | RBM + Drop out | SVM | **91.4** | | | **57.4** 77.4 (MCI:NC) | | | 51 | 43 | 56 | 52 | 202 |
| **Suk et al. (2015)** | MRI,PET,CSF | SAE + sparse learning | SVM | **98.8** | | | **83.3** 90.7 (MCI:NC) | | | 51 | 43 | 56 | 52 | 202 |
| **Liu et al. (2015)** | MRI,PET | SAE with zero-masking | softmax | **91.4** | 92.32 | 90.42 | 82.1 (MCI:NC) | 60.0 | 92.32 | 77 | 67 | 102 | 85 | 331 |
| **Cheng et al. (2017)** | MRI | 3D CNN | softmax | **87.15** | 86.36 | 85.93 | | | | 199 | | | 229 | 428 |
| **Cheng and Liu (2017)** | MRI,PET | 3D CNN + 2D CNN | softmax | **89.64** | 87.10 | 92.00 | | | | 93 | | | 100 | 193 |
| **Aderghal et al. (2017)** | MRI | 2D CNN | softmax | **91.41** | 93.75 | 89.06 | 65.62(MCI:MC) | 66.25 | 65.0 | 188 | 399 | | 228 | 815 |
| **Korolev et al. (2017)** | MRI | 3D CNN | softmax | **80** | 87 (AUC) | | 61(lMCI:NC) 56(eMCI:NC) | 65 (AUC) 58 (AUC) | | 50 | 43 (lMCI) | 77 (eMCI) | 61 | 111 |
| **Vu et al. (2017)** | MRI, PET | SAE+3D CNN | softmax | **91.14** | | | | | | 145 | | | 172 | 317 |
| **Liu et al. (2018a)** | PET | RNN | softmax | **91.2** | 91.4 | 91.0 | 78.9 (MCI:NC) | 78.1 | 80.0 | 93 | 146 | | 100 | 339 |
| **Liu et al. (2018b)** | MRI | Landmark detection + 3D CNN | softmax | **91.09** | 88.05 | 93.50 | **76.9** | 42.11 | 82.43 | 159 | 38 | 239 | 200 | 636 |
| **Lu et al. (2018)** | MRI, PET | DNN + NN | softmax | **84.6** | 80.2 | 91.8 | **82.93** | 79.69 | 83.84 | 238 | 217 | 409 | 360 | 1224 |
| **Choi and Jin (2018)** | PET | 3D CNN | softmax | **96** | 93.5 | 97.8 | **84.2** | 81.0 | 87.0 | 139 | 79 | 92 | 182 | 492 |

SEN = TP/ (TP + FN), SPE = TN/ (TN + FP)
(TP: true positive, TN: true negative, FP: false positive, FN: false negative)
All data on this table were from ADNI.

## 12  Figures

Figure 1. The multilayer perceptron procedure. After the initial error value is calculated from the given random weight by the least squares method, the weights are updated by a back-propagation algorithm until the differential value becomes 0.

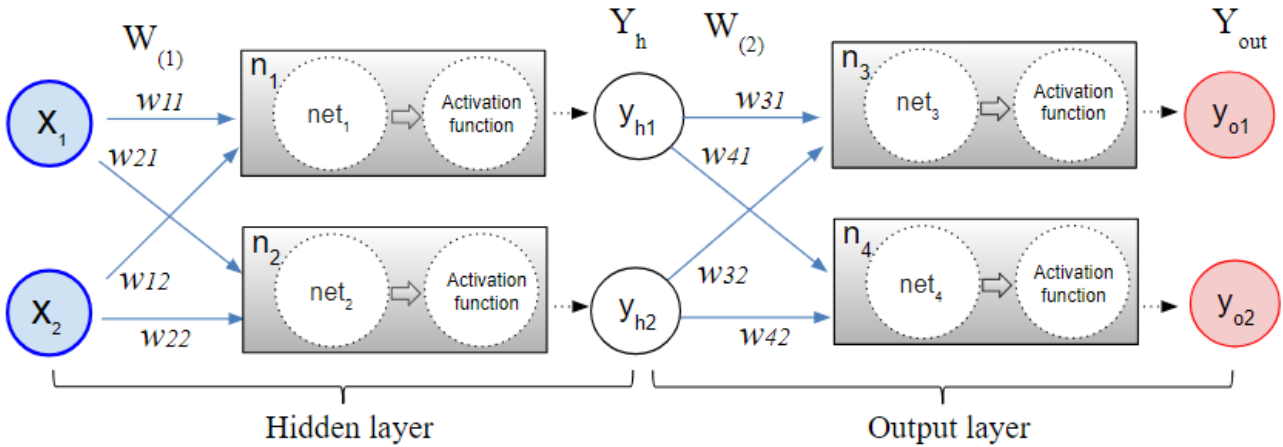

Figure 2. Common activation functions used in deep learning (red) and their derivatives (blue). When the sigmoid is differentiated, the maximum value is 0.25, which becomes closer to 0 when it continues to multiply.

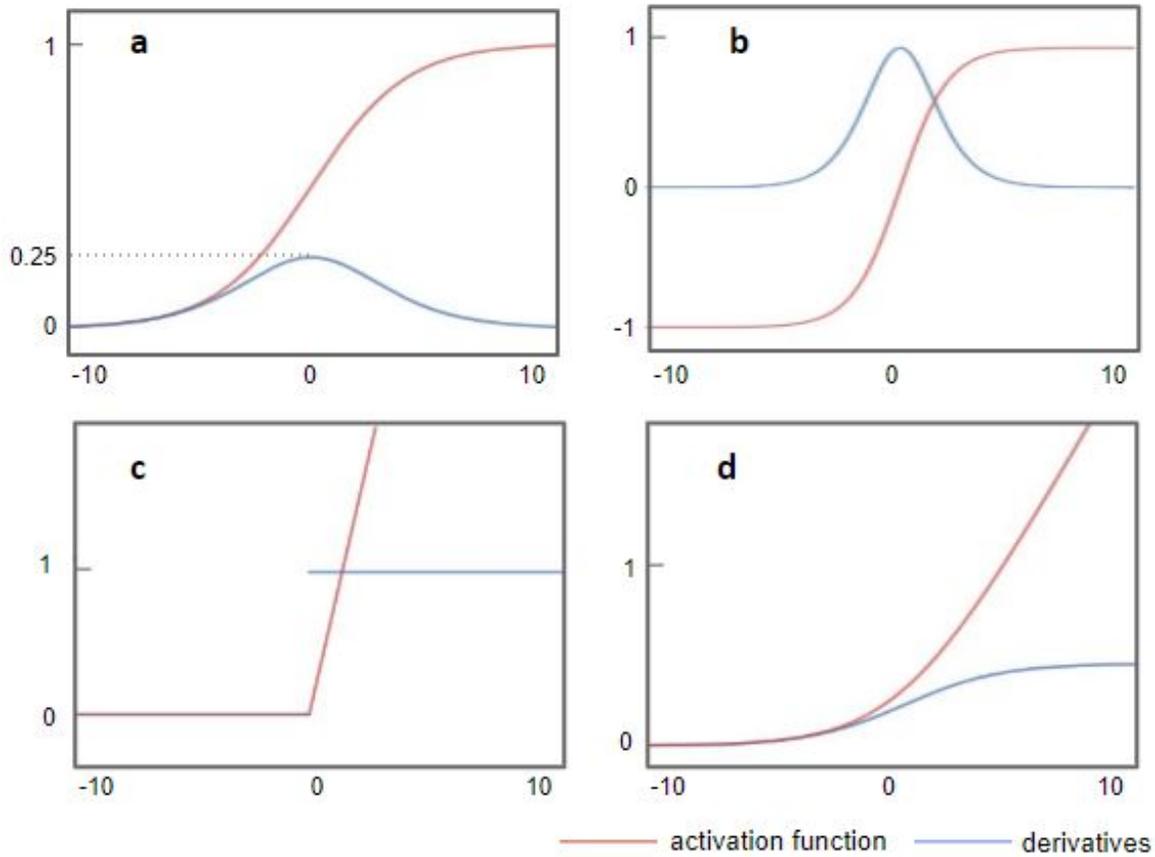

| a. sigmoid | b. tanh | c. relu | d. softplus |
| --- | --- | --- | --- |
| $f(x) = \frac{1}{1+e^{-x}}$ | $f(x) = \tanh(x)$ | $f(x) = \max(0, x)$ | $f(x) = \log(1 + e^x)$ |

Figure 3. Architectural structures in deep learning: (a) RBM (Hinton and Salakhutdinov, 2006) (b) DBM (Salakhutdinov and Larochelle, 2010) (c) DBN (Bengio, 2009) (d) CNN (Krizhevsky et al., 2012) (e) AE (Fukushima, 1975;Krizhevsky and Hinton, 2011) (f) Sparse AE (Vincent et al., 2008;Vincent et al., 2010) (g) Stacked AE (Larochelle et al., 2009;Makhzani and Frey, 2013). RBM=Restricted Boltzmann Machine; DBM=Deep Boltzmann Machine; DBN=Deep Belief Network; CNN=Convolutional Neural Network; AE=Auto-Encoders.

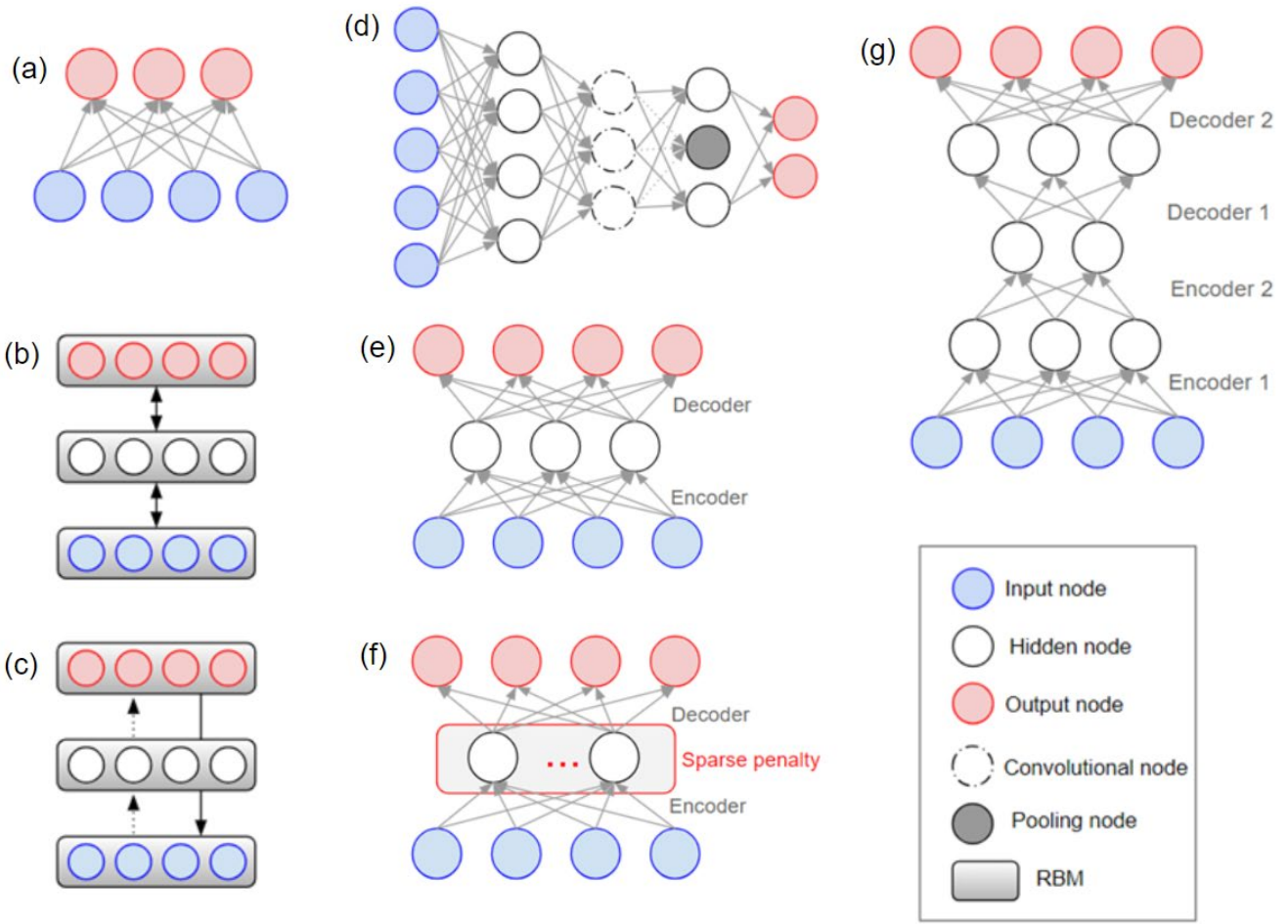

Figure 4. PRISMA (Preferred Reporting Items for Systematic Reviews and Meta-Analyses) Flow Chart. From a total of 389 hits on Google scholar and PubMed search, 16 articles were included in the systematic

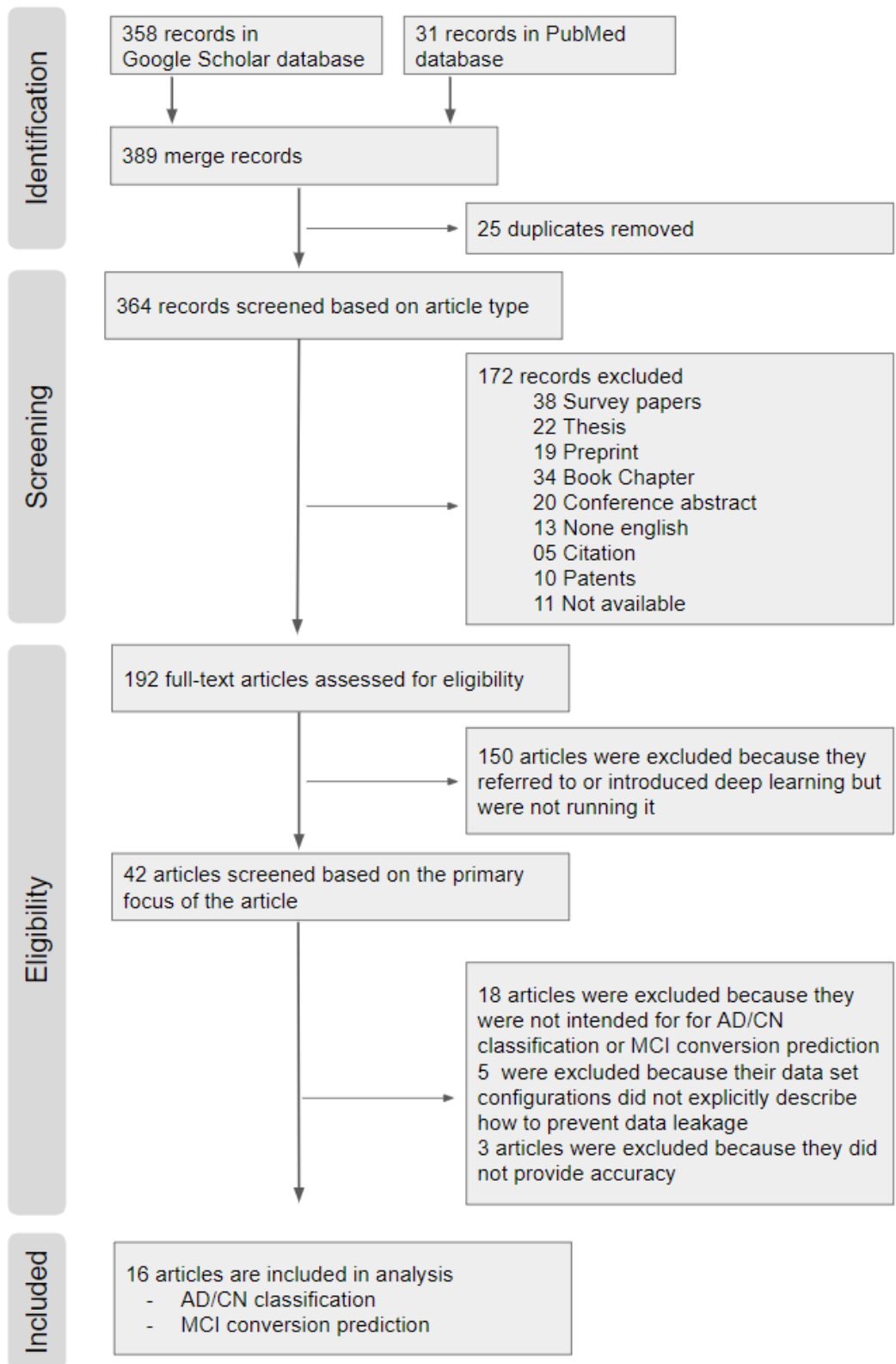

review.

Figure 5. Comparison of diagnostic classification accuracy of pure deep learning and hybrid approach. 4 studies (gray) have used hybrid methods that combine deep learning for feature selection from neuroimaging data and traditional machine learning, such as the SVM as a classifier. 12 studies (blue) have used deep learning method with softmax classifier for diagnostic classification and/or prediction of MCI to AD conversion.

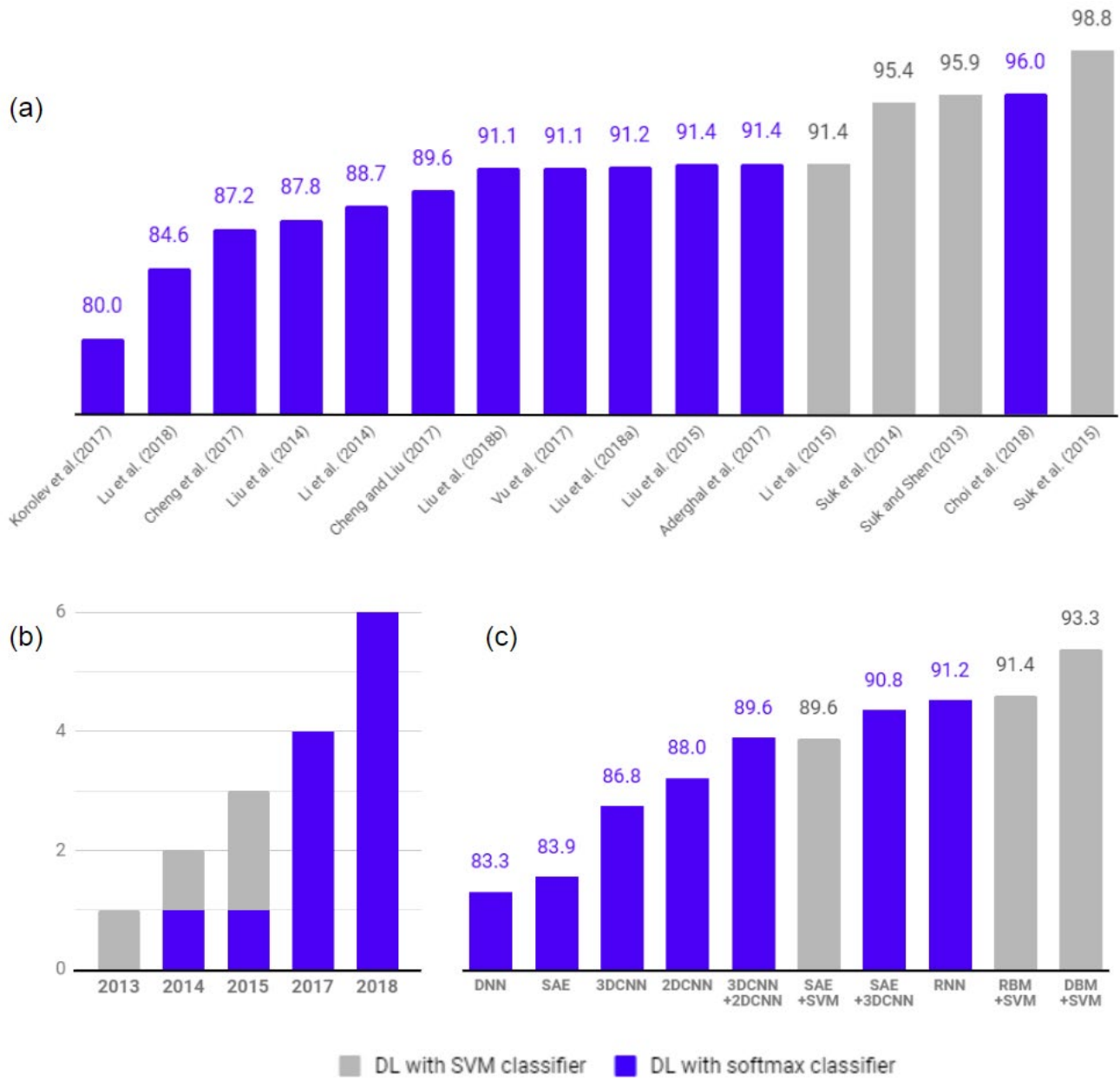

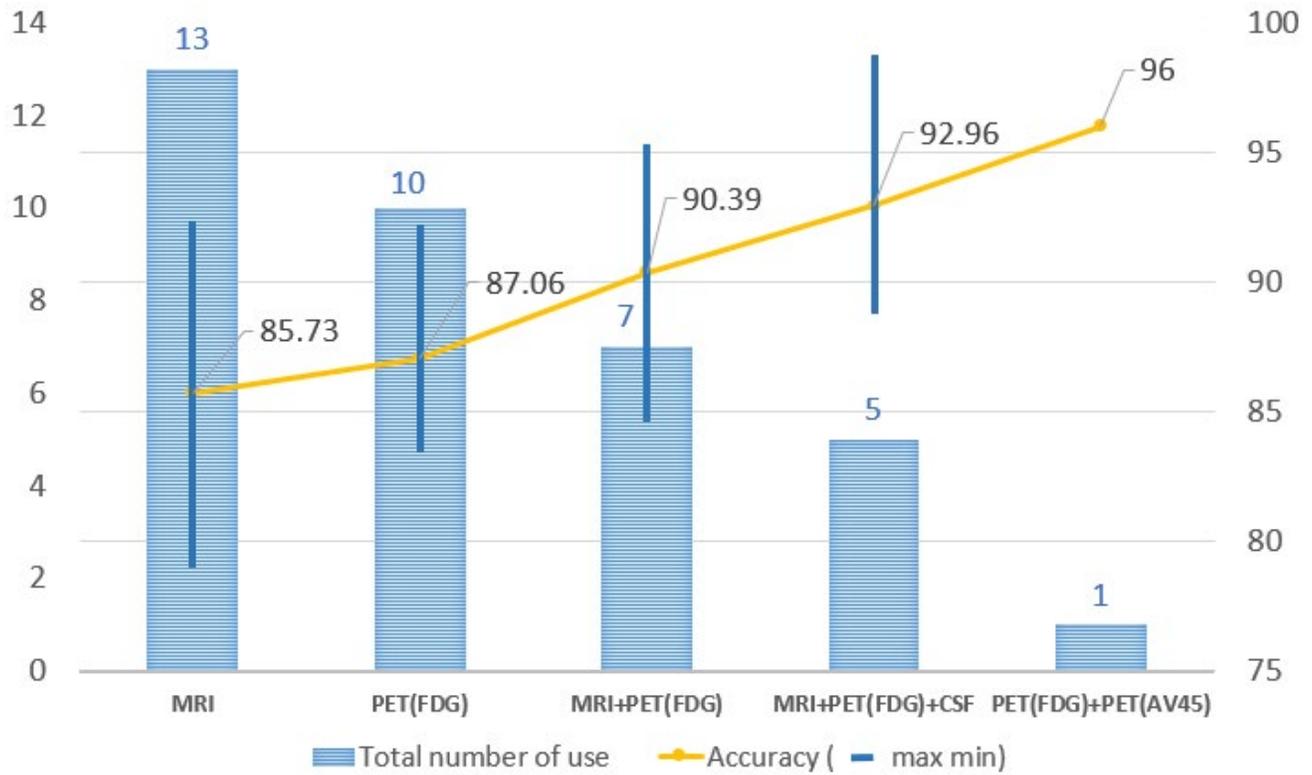

Figure 6. Changes in accuracy by types of image resource. MRI scans were used in 13 studies, FDG-PET scans in 10, both MRI and FDG-PET scans in 12, and both amyloid PET and FDG-PET scans in 1. The performance in AD/CN classification yielded better results in PET data compared to MRI. Two or more multimodal neuroimaging data types produced higher accuracies than a single neuroimaging technique.

# 13    Reference

Aderghal, K., Benois-Pineau, J., Afdel, K., and Catheline, G. (2017). "FuseMe: Classification of sMRI images by fusion of Deep CNNs in 2D+ϵ projections", in: *Proceedings of the 15th International Workshop on Content-Based Multimedia Indexing.*

Bengio, Y. (2009). Learning deep architectures for AI. *Foundations and trends in Machine Learning* 2**,** 1-127.

Bengio, Y. (2013). "Deep learning of representations: Looking forward", in: *International Conference on Statistical Language and Speech Processing*: Springer), 1-37.

Bengio, Y., Courville, A., and Vincent, P. (2013). Representation learning: A review and new perspectives. *IEEE transactions on pattern analysis and machine intelligence* 35**,** 1798-1828.

Bishop, C.M. (1995). *Neural networks for pattern recognition.* Oxford university press.

Bottou, L. (2010). "Large-scale machine learning with stochastic gradient descent," in *Proceedings of COMPSTAT'2010.* Springer), 177-186.

Boureau, Y.-L., Ponce, J., and Lecun, Y. (2010). "A theoretical analysis of feature pooling in visual recognition", in: *Proceedings of the 27th international conference on machine learning (ICML-10)*), 111-118.

Bush, W.S., and Moore, J.H. (2012). Genome-wide association studies. *PLoS computational biology* 8**,** e1002822.

Cheng, D., and Liu, M. (2017). "CNNs based multi-modality classification for AD diagnosis", in: *2017 10th International Congress on Image and Signal Processing, BioMedical Engineering and Informatics (CISP-BMEI)*), 1-5.

Cheng, D., Liu, M., Fu, J., and Wang, Y. (2017). "Classification of MR brain images by combination of multi-CNNs for AD diagnosis", in: *Ninth International Conference on Digital Image Processing (ICDIP 2017)*: SPIE), 5.

Choi, H., and Jin, K.H. (2018). Predicting cognitive decline with deep learning of brain metabolism and amyloid imaging. *Behavioural Brain Research* 344**,** 103-109.

Ciregan, D., Meier, U., and Schmidhuber, J. (2012). "Multi-column deep neural networks for image classification", in: *2012 IEEE Conference on Computer Vision and Pattern Recognition*), 3642-3649.

De strooper, B., and Karran, E. (2016). The Cellular Phase of Alzheimer's Disease. *Cell* 164**,** 603-615.

Farabet, C., Couprie, C., Najman, L., and Lecun, Y. (2013). Learning Hierarchical Features for Scene Labeling. *IEEE Transactions on Pattern Analysis and Machine Intelligence* 35**,** 1915-1929.

Fukushima, K. (1979). Neural network model for a mechanism of pattern recognition unaffected by shift in position-Neocognitron. *IEICE Technical Report, A* 62**,** 658-665.

Fukushima, K. (1980). Neocognitron: A self-organizing neural network model for a mechanism of pattern recognition unaffected by shift in position. *Biological cybernetics* 36**,** 193-202.

Galvin, J.E. (2017). Prevention of Alzheimer's Disease: Lessons Learned and Applied. *J Am Geriatr Soc* 65**,** 2128-2133.

Glorot, X., Bordes, A., and Bengio, Y. (2011). "Deep sparse rectifier neural networks", in: *Proceedings of the fourteenth international conference on artificial intelligence and statistics*), 315-323.

Goodfellow, I., Bengio, Y., Courville, A., and Bengio, Y. (2016). *Deep learning.* MIT press Cambridge.

Goodfellow, I., Pouget-Abadie, J., Mirza, M., Xu, B., Warde-Farley, D., Ozair, S., Courville, A., and Bengio, Y. (2014). "Generative adversarial nets", in: *Advances in neural information processing systems*), 2672-2680.

Gulshan, V., Peng, L., Coram, M., and Et Al. (2016). Development and validation of a deep learning algorithm for detection of diabetic retinopathy in retinal fundus photographs. *JAMA* 316**,** 2402-2410.

Hinton, G., Deng, L., Yu, D., Dahl, G., Mohamed, A.-R., Jaitly, N., Senior, A., Vanhoucke, V., Nguyen, P., Kingsbury, B., and Sainath, T. (2012). Deep Neural Networks for Acoustic Modeling in Speech Recognition. *IEEE Signal Processing Magazine* 29**,** 82-97.

Hinton, G.E., Osindero, S., and Teh, Y.-W. (2006). A fast learning algorithm for deep belief nets. *Neural computation* 18**,** 1527-1554.

Hinton, G.E., and Salakhutdinov, R.R. (2006). Reducing the dimensionality of data with neural networks. *science* 313**,** 504-507.

Hinton, G.E., and Zemel, R.S. (1994). "Autoencoders, minimum description length and Helmholtz free energy", in: *Advances in neural information processing systems*), 3-10.

Ivakhnenko, A.G. (1968). The Group Method of Data of Handling; A rival of the method of stochastic approximation. *Soviet Automatic Control* 13**,** 43-55.

Ivakhnenko, A.G. (1971). Polynomial theory of complex systems. *IEEE transactions on Systems, Man, and Cybernetics***,** 364-378.

Ivakhnenko, A.G.E., and Lapa, V.G. (1965). *Cybernetic predicting devices.* CCM Information Corporation.

König, I.R. (2011). Validation in Genetic Association Studies. *Briefings in Bioinformatics* 12**,** 253-258.

Kononenko, I. (2001). Machine learning for medical diagnosis: history, state of the art and perspective. *Artificial Intelligence in medicine* 23**,** 89-109.

Korolev, S., Safiullin, A., Belyaev, M., and Dodonova, Y. (2017). "Residual and plain convolutional neural networks for 3D brain MRI classification", in: *2017 IEEE 14th International Symposium on Biomedical Imaging (ISBI 2017)*), 835-838.

Krizhevsky, A., Sutskever, I., and Hinton, G.E. (2012). "Imagenet classification with deep convolutional neural networks", in: *Advances in neural information processing systems*), 1097-1105.

Lecun, Y., Bengio, Y., and Hinton, G. (2015). Deep learning. *Nature* 521**,** 436.

Lecun, Y., Touresky, D., Hinton, G., and Sejnowski, T. (1988). "A theoretical framework for back-propagation", in: *Proceedings of the 1988 connectionist models summer school*: CMU, Pittsburgh, Pa: Morgan Kaufmann), 21-28.

Li, F., Tran, L., Thung, K.-H., Ji, S., Shen, D., and Li, J. (2015). A Robust Deep Model for Improved Classification of AD/MCI Patients. *IEEE journal of biomedical and health informatics* 19**,** 1610-1616.

Li, R., Zhang, W., Suk, H.-I., Wang, L., Li, J., Shen, D., and Ji, S. (2014). Deep learning based imaging data completion for improved brain disease diagnosis. *Medical image computing and computer-assisted intervention : MICCAI ... International Conference on Medical Image Computing and Computer-Assisted Intervention* 17**,** 305-312.

Litjens, G., Kooi, T., Bejnordi, B.E., Setio, A.a.A., Ciompi, F., Ghafoorian, M., Van Der Laak, J., Van Ginneken, B., and Sanchez, C.I. (2017). A survey on deep learning in medical image analysis. *Med Image Anal* 42**,** 60-88.

Liu, M., Cheng, D., Yan, W., and , A.S.D.N.I. (2018a). Classification of Alzheimer's Disease by Combination of Convolutional and Recurrent Neural Networks Using FDG-PET Images. *Frontiers in Neuroinformatics* 12.

Liu, M., Zhang, J., Adeli, E., and Shen, D. (2018b). Landmark-based deep multi-instance learning for brain disease diagnosis. *Medical Image Analysis* 43**,** 157-168.

Liu, S., Liu, S., Cai, W., Che, H., Pujol, S., Kikinis, R., Feng, D., Fulham, M.J., and Adni (2015). Multimodal Neuroimaging Feature Learning for Multiclass Diagnosis of Alzheimer's Disease. *IEEE Transactions on Biomedical Engineering* 62**,** 1132-1140.

Liu, S., Liu, S., Cai, W., Pujol, S., Kikinis, R., and Feng, D. (2014). "Early diagnosis of Alzheimer's disease with deep learning", in: *2014 IEEE 11th International Symposium on Biomedical Imaging (ISBI)*), 1015-1018.

Lu, D., Popuri, K., Ding, G.W., Balachandar, R., and Beg, M.F. (2018). Multimodal and Multiscale Deep Neural Networks for the Early Diagnosis of Alzheimer's Disease using structural MR and FDG-PET images. *Scientific Reports* 8**,** 5697.

Lu, D., and Weng, Q. (2007). A survey of image classification methods and techniques for improving classification performance. *International journal of Remote sensing* 28**,** 823-870.

Makhzani, A., and Frey, B. (2013). K-sparse autoencoders. *arXiv preprint arXiv:1312.5663*.

Marcus, G. (2018). Deep learning: A critical appraisal. *arXiv preprint arXiv:1801.00631*.

Mcculloch, W.S., and Pitts, W. (1943). A logical calculus of the ideas immanent in nervous activity. *The bulletin of mathematical biophysics* 5**,** 115-133.

Mikolov, T., Sutskever, I., Chen, K., Corrado, G.S., and Dean, J. (2013). Distributed Representations of Words and Phrases and their Compositionality. 3111--3119.

Minsky, M., and Papert, S. (1969). Perceptrons. *Cambridge, MA: MIT Press* 18**,** 19.

Moher, D., Liberati, A., Tetzlaff, J., and Altman, D.G. (2009). Preferred reporting items for systematic reviews and meta-analyses: the PRISMA statement. *Annals of internal medicine* 151**,** 264-269.

Nair, V., and Hinton, G.E. (2010). "Rectified linear units improve restricted boltzmann machines", in: *Proceedings of the 27th international conference on machine learning (ICML-10)*), 807-814.

Ngiam, J., Khosla, A., Kim, M., Nam, J., Lee, H., and Ng, A.Y. (2011). "Multimodal deep learning", in: *Proceedings of the 28th international conference on machine learning (ICML-11)*), 689-696.

Plis, S.M., Hjelm, D.R., Salakhutdinov, R., Allen, E.A., Bockholt, H.J., Long, J.D., Johnson, H.J., Paulsen, J.S., Turner, J.A., and Calhoun, V.D. (2014). Deep learning for neuroimaging: a validation study. *Frontiers in Neuroscience* 8.

Rathore, S., Habes, M., Iftikhar, M.A., Shacklett, A., and Davatzikos, C. (2017). A review on neuroimaging-based classification studies and associated feature extraction methods for Alzheimer's disease and its prodromal stages. *NeuroImage* 155**,** 530-548.

Reas, E. (2018). ADNI: Understanding Alzheimer's disease through collaboration and data sharing. *PLOS blogs*.

Riedel, B.C., Daianu, M., Ver Steeg, G., Mezher, A., Salminen, L.E., Galstyan, A., and Thompson, P.M. (2018). Uncovering Biologically Coherent Peripheral Signatures of Health and Risk for Alzheimer's Disease in the Aging Brain. *Front Aging Neurosci* 10**,** 390.

Ripley, B.D., and Hjort, N. (1996). *Pattern recognition and neural networks.* Cambridge university press.

Rosenblatt, F. (1957). *The perceptron, a perceiving and recognizing automaton Project Para.* Cornell Aeronautical Laboratory.

Rosenblatt, F. (1958). The perceptron: a probabilistic model for information storage and organization in the brain. *Psychological review* 65**,** 386.

Rumelhart, D.E., Hinton, G.E., and Williams, R.J. (1986). Learning representations by back-propagating errors. *nature* 323**,** 533.

Russakovsky, O., Deng, J., Su, H., Krause, J., Satheesh, S., Ma, S., Huang, Z., Karpathy, A., Khosla, A., and Bernstein, M. (2015). Imagenet large scale visual recognition challenge. *International journal of computer vision* 115**,** 211-252.

Salakhutdinov, R., and Larochelle, H. (2010). "Efficient learning of deep Boltzmann machines", in: *Proceedings of the thirteenth international conference on artificial intelligence and statistics*), 693-700.

Samper-Gonzalez, J., Burgos, N., Bottani, S., Fontanella, S., Lu, P., Marcoux, A., Routier, A., Guillon, J., Bacci, M., Wen, J., Bertrand, A., Bertin, H., Habert, M.O., Durrleman, S., Evgeniou, T., and Colliot, O. (2018). Reproducible evaluation of classification methods in Alzheimer's disease: Framework and application to MRI and PET data. *Neuroimage* 183**,** 504-521.

Schalkoff, R.J. (1997). *Artificial neural networks.* McGraw-Hill New York.

Schelke, M.W., Attia, P., Palenchar, D.J., Kaplan, B., Mureb, M., Ganzer, C.A., Scheyer, O., Rahman, A., Kachko, R., Krikorian, R., Mosconi, L., and Isaacson, R.S. (2018). Mechanisms of Risk Reduction in the Clinical Practice of Alzheimer's Disease Prevention. *Front Aging Neurosci* 10**,** 96.

Schmidhuber, J. (2015). Deep learning in neural networks: An overview. *Neural networks* 61**,** 85-117.

Smialowski, P., Frishman, D., and Kramer, S. (2009). Pitfalls of supervised feature selection. *Bioinformatics* 26**,** 440-443.

Suk, H.-I., Lee, S.-W., Shen, D., and The Alzheimer's Disease Neuroimaging, I. (2015). Latent feature representation with stacked auto-encoder for AD/MCI diagnosis. *Brain structure & function* 220**,** 841-859.

Suk, H.-I., Lee, S.-W., Shen, D., and The Alzheimers Disease Neuroimaging, I. (2014). Hierarchical Feature Representation and Multimodal Fusion with Deep Learning for AD/MCI Diagnosis. *NeuroImage* 101**,** 569-582.

Suk, H.-I., and Shen, D. (2013). Deep Learning-Based Feature Representation for AD/MCI Classification. *Medical image computing and computer-assisted intervention : MICCAI ... International Conference on Medical Image Computing and Computer-Assisted Intervention* 16**,** 583-590.

Sutskever, I., Martens, J., Dahl, G., and Hinton, G. (2013). "On the importance of initialization and momentum in deep learning", in: *International conference on machine learning*), 1139-1147.

Sutton, R.S., and Barto, A.G. (2018). *Reinforcement learning: An introduction.* MIT press.

Toga, A.W., Bhatt, P., and Ashish, N. (2016). Global data sharing in Alzheimer's disease research. *Alzheimer disease and associated disorders* 30**,** 160.

Veitch, D.P., Weiner, M.W., Aisen, P.S., Beckett, L.A., Cairns, N.J., Green, R.C., Harvey, D., Jack, C.R., Jr., Jagust, W., Morris, J.C., Petersen, R.C., Saykin, A.J., Shaw, L.M., Toga, A.W., and Trojanowski, J.Q. (2019). Understanding disease progression and improving Alzheimer's disease clinical trials: Recent highlights from the Alzheimer's Disease Neuroimaging Initiative. *Alzheimers Dement* 15**,** 106-152.

Vu, T.D., Yang, H.-J., Nguyen, V.Q., Oh, A.R., and Kim, M.-S. (2017). "Multimodal learning using convolution neural network and Sparse Autoencoder", in: *2017 IEEE International Conference on Big Data and Smart Computing (BigComp)*), 309-312.

Werbos, P.J. (1982). "Applications of advances in nonlinear sensitivity analysis," in *System modeling and optimization*. Springer), 762-770.

Werbos, P.J. (2006). "Backwards differentiation in AD and neural nets: Past links and new opportunities," in *Automatic differentiation: Applications, theory, and implementations*. Springer), 15-34.

*Supplementary Material*

# Deep Learning in Alzheimer's Disease: Diagnostic Classification and Prognostic Prediction using Neuroimaging Data

**Taeho Jo[1-3]\*, Kwangsik Nho[1-3], Andrew J. Saykin[1-3]**

[1]Center for Neuroimaging, Department of Radiology and Imaging Sciences, Indiana University School of Medicine, Indianapolis, IN, USA [2]Indiana Alzheimer Disease Center, Indiana University School of Medicine, Indianapolis, IN, USA

[3]Indiana University Network Science Institute, Bloomington, IN, USA

**\* Correspondence:** Corresponding Author tjo@iu.edu

**Supplement 1. Weights calculation in the backpropagation**

After the initial error value is calculated from the given random weight by the least squares method, the weights are updated until the differential value becomes 0. The differential value 0 means there is no change in weight when the gradient is subtracted from the previous weight. In Fig. 1, the $w_{31}$ is updated by following formula:

$$w_{31}(t+1) = w_{31}t - \frac{\partial ErrorY_{out}}{\partial w_{31}}$$

$$ErrorY_{out} = \frac{1}{2}(y_{t1} - y_{o1})^2 + \frac{1}{2}(y_{t2} - y_{o2})^2$$

The $ErrorY_{out}$ is the sum of error $y_{o1}$ and error $y_{o2}$. $y_{t1}$, $y_{t2}$ are constants that are known through the given data. The partial derivative of $ErrorY_{out}$ with respect to $w_{31}$ can be calculated by the chain rule as follows.

$$\frac{\partial ErrorY_{out}}{\partial w_{31}} = \frac{\partial ErrorY_{out}}{\partial y_{o1}} \cdot \frac{\partial y_{o1}}{\partial net3} \cdot \frac{\partial net3}{\partial w_{31}}$$

$$(i) \qquad (ii) \qquad (iii)$$

Here, (i) becomes $y_{o1}$-$y_{t1}$ which is the partial derivative of $\frac{1}{2}(y_{t1} - y_{o1})^2$ with respect to $y_{o1}$. When activation function $\sigma(x)$ is $\frac{1}{1+e^{-x}}$, $\frac{d\sigma(x)}{dx} = \sigma(x) \cdot (1-\sigma(x))$ which makes (ii) $y_{o1} \cdot (1-y_{o1})$. Since $Net_3$ is $w_{31}y_{h1} + w_{41}y_{h2}$ + bias, the partial derivative of $Net_3$ with respect to $w_{31}$, which (iii), is $y_{h1}$.

$$w_{31}(t+1) = w_{31}t - (y_{o1} - y_{t1})y_{o1}(1 - y_{o1})y_{h1}$$

To update $W_{11}$ in hidden layer, it is also started from $ErrorY_{out}$, since $Y_h$ is located in the hidden layer and is not exposed.

$$w_{11}(t+1) = w_{11}t - \frac{\partial ErrorY_{out}}{\partial w_{11}}$$

$$\frac{\partial ErrorY_{out}}{\partial w_{11}} = \frac{\partial ErrorY_{out}}{\partial y_{h1}} \cdot \frac{\partial y_{h1}}{\partial net_1} \cdot \frac{\partial net_1}{\partial w_{11}}$$

(i)

Here, the calculation of (i) is a bit different from previous. Since $ErrorY_{out}$ includes $Error_{yo1}$ and $Error_{yo2}$, it is calculated as follows.

$$\frac{\partial ErrorY_{out}}{\partial h_1} = \frac{\partial(Error_{yo1} + Error_{yo2})}{\partial y_{h1}} = \frac{\partial Error_{yo1}}{\partial y_{h1}} + \frac{\partial Error_{yo2}}{\partial y_{h1}}$$

(a)      (b)

(a), (b) is calculated as follows by the chain rule.

(a) $\frac{\partial Error_{yo1}}{\partial y_{h1}} = \frac{\partial Error_{yo1}}{\partial net_3} \cdot \frac{\partial net_3}{\partial y_{h1}} = (y_{o1} - y_{t1})y_{o1}(1 - y_{o1})y_{o1}$

(b) $\frac{\partial Error_{yo2}}{\partial y_{h1}} = \frac{\partial Error_{yo2}}{\partial net_4} \cdot \frac{\partial net_4}{\partial y_{h1}} = (y_{o2} - y_{t2})y_{o2}(1 - y_{o2})y_{o2}$

Now, (i), (ii), and (iii) are summarized as follows.

$$\frac{\partial ErrorY_{out}}{\partial w_{11}} = \frac{\partial ErrorY_{out}}{\partial y_{h1}} \cdot \frac{\partial y_{h1}}{\partial net_1 y} \cdot \frac{\partial net_1}{\partial w_{11}}$$

$$= (\delta y_{o1} y_{o1} - \delta y_{o2} y_{o2}) y_{h1}(1 - y_{h1}) x_1$$

**Supplement 2. Advanced gradient descent methods**

Nesterov Momentum is the method of adding the value of $\gamma v_{(t-1)}$ to the Momentum SGD to find the gradient. This allows to reduce unnecessary movements by advance movement in the direction to move.

$$w_{(t+1)} = w_t + \gamma v_{(t-1)} - \eta \frac{\partial Error}{\partial (w + \gamma v_{(t-1)})}$$

Adagrad(Adaptive Gradient) is an optimization method that adjusts the learning rate according to the number of update of variables.

$$G_t = G_{(t-1)} + \left(\frac{\partial Error}{\partial w_t}\right)^2$$

$$w_{(t+1)} = w_t - \eta \frac{1}{\sqrt{G_t + \epsilon}} \frac{\partial Error}{\partial w_t}$$

Here, RMSprop is the method of adjusting the ratio between the previous value and the modified value.

$$G_t = \gamma G_{(t-1)} + (1 - \gamma)\left(\frac{\partial Error}{\partial w_t}\right)^2$$

$$w_{(t+1)} = w_t - \eta \frac{1}{\sqrt{G_t + \epsilon}} \frac{\partial Error}{\partial w_t}$$

Adam, the most popular optimization method for deep learning today, takes advantage of momentum SGD and RMSprop. Adam is expressed as follows. Where $G_t$ is the sum of the square of the modified gradient. $\epsilon$ is a very small constant that prevents it from being divided by zero.

$$V_t = \gamma G_{(t-1)} + (1-\gamma_1)\frac{\partial Error}{\partial w_t}$$

$$G_t = \gamma G_{(t-1)} + (1-\gamma_2)\left(\frac{\partial Error}{\partial w_t}\right)^2$$

$$\widehat{V}_t = \frac{V_t}{1-\gamma_1^t} \quad \widehat{G}_t = \frac{G_t}{1-\gamma_2^t}$$

$$w_{(t+1)} = w_t - \eta \frac{\widehat{G}_t}{\sqrt{\widehat{V}_t} + \epsilon}$$

**Supplement 3.** Table S1. All the results of the 16 studies to systematically be reviewed

| Author (year) | Modality | Data processing, training | Classifier | AD | cMCI | ncMCI | NC | Total | Acc. AD/NC | STD | Acc. MCI conversion | STD |
|---|---|---|---|---|---|---|---|---|---|---|---|---|
| Suk et al. (2015) | MRI,PET,CSF | SAE + sparse learning | SVM | 51 | 43 | 56 | 52 | 202 | **98.8** | 0.4 | **83.3** | 2.1 |
| Choi and Jin (2018) | PET | 3D CNN | softmax | 139 | 79 | 92 | 182 | 492 | **96** | | **84.2** | |
| Suk and Shen (2013) | MRI,PET,CSF | SAE | SVM | 51 | 43 | 56 | 52 | 202 | **95.9** | 1.1 | **75.8** | 2 |
| Suk et al. (2014) | MRI,PET | DBM | SVM | 93 | 76 | 128 | 101 | 398 | **95.35** | 5.23 | **75.92** | 15.37 |
| Li et al. (2014) | MRI, PET | 3D CNN | Logistic regression | 198 | 167 | 236 | 229 | 830 | **92.87** | 2.07 | **72.44** | 2.41 |
| Suk et al. (2015) | MRI | SAE + sparse learning | SVM | 51 | 43 | 56 | 52 | 202 | **92.4** | 1.5 | **69.3** | 2 |
| Suk et al. (2014) | MRI | DBM | SVM | 93 | 76 | 128 | 101 | 398 | **92.38** | 5.32 | **72.42** | 13.09 |
| Suk et al. (2014) | PET | DBM | SVM | 93 | 76 | 128 | 101 | 398 | **92.2** | 6.7 | **70.25** | 13.23 |
| Li et al. (2014) | MRI | 3D CNN | Logistic regression | 198 | 167 | 236 | 229 | 830 | **91.92** | 1.88 | **71.68** | 2.53 |
| Aderghal et al. (2017) | MRI | 2D CNN | softmax | 188 | 399 | | 228 | 815 | **91.41** | | | |
| Li et al. (2015) | MRI,PET,CSF | RBM + Drop out | SVM | 51 | 43 | 56 | 52 | 202 | **91.4** | 1.8 | **57.4** | 3.6 |
| Liu et al. (2015) | MRI,PET | SAE with zero-masking | softmax | 77 | 67 | 102 | 85 | 331 | **91.4** | 5.56 | | |
| Liu et al. (2018a) | PET | RNN | softmax | 93 | 146 | | 100 | 339 | **91.2** | | | |
| Vu et al. (2017) | MRI, PET | SAE + 3D CNN | softmax | 145 | | | 172 | 317 | **91.14** | | | |
| Liu et al. (2018b) | MRI | Landmark detection + 3D CNN | softmax | 159 | 38 | 239 | 200 | 636 | **91.09** | | **76.9** | |
| Cheng and Liu (2017) | MRI,PET | 3D CNN + 2D CNN | softmax | 93 | | | 100 | 193 | **89.64** | | | |
| Suk et al. (2015) | PET | SAE + sparse learning | SVM | 51 | 43 | 56 | 52 | 202 | **88.7** | 2.7 | **68.9** | 3.8 |
| Li et al. (2014) | PET | 3D CNN | Logistic regression | 198 | 167 | 236 | 229 | 830 | **87.62** | 2.36 | **70.29** | 2.45 |
| Liu et al. (2014) | MRI,PET | SAE+NN | softmax | 65 | 67 | 102 | 77 | 311 | **87.76** | | | |
| Cheng et al. (2017) | MRI | 3D CNN | softmax | 199 | | | 229 | 428 | **87.15** | | | |
| Cheng and Liu (2017) | PET | 3D CNN + 2D CNN | softmax | 93 | | | 100 | 193 | **87.13** | | | |
| Cheng and Liu (2017) | MRI | 3D CNN + 2D CNN | softmax | 93 | | | 100 | 193 | **85.47** | | | |
| Lu et al. (2018) | MRI,PET | DNN + NN | softmax | 238 | 217 | 409 | 360 | 1224 | **84.6** | 1.5 | **82.93** | 7.25 |
| Lu et al. (2018) | PET | DNN + NN | softmax | 238 | 217 | 409 | 360 | 1224 | **84.5** | 1.4 | **81.53** | 7.42 |
| Lu et al. (2018) | MRI | DNN + NN | softmax | 238 | 217 | 409 | 360 | 1224 | **81.9** | 1.2 | **75.44** | 7.74 |
| Korolev et al. (2017) | MRI | 3D CNN | softmax | 50 | | | 61 | 111 | **80** | 7 | | |
| Suk et al. (2015) | CSF | SAE + sparse learning | SVM | 51 | 43 | 56 | 52 | 202 | **79.7** | 1.4 | **57.7** | 3 |

All data on this table were from ADNI.
https://github.com/rasmusbergpalm/DeepLearnToolbox (Suk et al. (2015), Suk and Shen (2013))
https://github.com/neuro-ml/resnet_cnn_mri_adni (Korolev et al. (2017))

**Reference**

Aderghal, K., Benois-Pineau, J., Afdel, K., and Catheline, G. (2017). "FuseMe: Classification of sMRI images by fusion of Deep CNNs in 2D+E projections", in: *Proceedings of the 15th International Workshop on Content-Based Multimedia Indexing*.

Cheng, D., Liu, M., Fu, J., and Wang, Y. (2017). "Classification of MR brain images by combination of multi-CNNs for AD diagnosis", in: *Ninth International Conference on Digital Image Processing (ICDIP 2017)*: SPIE), 5.

Cheng, D., and Liu, M. (2017). "CNNs based multi-modality classification for AD diagnosis", in: *2017 10th International Congress on Image and Signal Processing, BioMedical Engineering and Informatics (CISP-BMEI)*), 1-5.

Choi, H., and Jin, K.H. (2018). Predicting cognitive decline with deep learning of brain metabolism and amyloid imaging. *Behavioural Brain Research* 344**,** 103-109.

Korolev, S., Safiullin, A., Belyaev, M., and Dodonova, Y. (2017). "Residual and plain convolutional neural networks for 3D brain MRI classification", in: *2017 IEEE 14th International Symposium on Biomedical Imaging (ISBI 2017)*), 835-838.

Li, R., Zhang, W., Suk, H.-I., Wang, L., Li, J., Shen, D., and Ji, S. (2014). Deep learning based imaging data completion for improved brain disease diagnosis. *Medical image computing and computer-assisted intervention : MICCAI ... International Conference on Medical Image Computing and Computer-Assisted Intervention* 17**,** 305-312.

Li, F., Tran, L., Thung, K.-H., Ji, S., Shen, D., and Li, J. (2015). A Robust Deep Model for Improved Classification of AD/MCI Patients. *IEEE journal of biomedical and health informatics* 19**,** 1610-1616.

Liu, M., Cheng, D., Yan, W., and , A.S.D.N.I. (2018a). Classification of Alzheimer's Disease by Combination of Convolutional and Recurrent Neural Networks Using FDG-PET Images. *Frontiers in Neuroinformatics* 12.

Liu, S., Liu, S., Cai, W., Pujol, S., Kikinis, R., and Feng, D. (2014). "Early diagnosis of Alzheimer's disease with deep learning", in: *2014 IEEE 11th International Symposium on Biomedical Imaging (ISBI)*), 1015-1018.

Liu, S., Liu, S., Cai, W., Che, H., Pujol, S., Kikinis, R., Feng, D., Fulham, M.J., and Adni (2015). Multimodal Neuroimaging Feature Learning for Multiclass Diagnosis of Alzheimer's Disease. *IEEE Transactions on Biomedical Engineering* 62**,** 1132-1140.

Liu, M., Zhang, J., Adeli, E., and Shen, D. (2018b). Landmark-based deep multi-instance learning for brain disease diagnosis. *Medical Image Analysis* 43**,** 157-168.

Lu, D., Popuri, K., Ding, G.W., Balachandar, R., and Beg, M.F. (2018). Multimodal and Multiscale Deep Neural Networks for the Early Diagnosis of Alzheimer's Disease using structural MR and FDG-PET images. *Scientific Reports* 8**,** 5697.

Suk, H.-I., and Shen, D. (2013). Deep Learning-Based Feature Representation for AD/MCI Classification. *Medical image computing and computer-assisted intervention : MICCAI ... International Conference on Medical Image Computing and Computer-Assisted Intervention* 16**,** 583-590.

Suk, H.-I., Lee, S.-W., Shen, D., and The Alzheimers Disease Neuroimaging, I. (2014). Hierarchical Feature Representation and Multimodal Fusion with Deep Learning for AD/MCI Diagnosis. *NeuroImage* 101**,** 569-582.

Suk, H.-I., Lee, S.-W., Shen, D., and The Alzheimer's Disease Neuroimaging, I. (2015). Latent feature representation with stacked auto-encoder for AD/MCI diagnosis. *Brain structure & function* 220**,** 841-859.

Vu, T.D., Yang, H.-J., Nguyen, V.Q., Oh, A.R., and Kim, M.-S. (2017). "Multimodal learning using convolution neural network and Sparse Autoencoder", in: *2017 IEEE International Conference on Big Data and Smart Computing (BigComp)*), 309-312.32